\documentclass[12pt]{article}
\title{The integration of systems of linear PDEs using conservation
       laws of syzygies}
\author{Thomas Wolf\\ 
Department of Mathematics \\
Brock University, 500 Glenridge Avenue,\\ 
St.Catharines, Ontario, Canada L2S 3A1 \\
email: twolf@brocku.ca}
\begin{document}
\maketitle
\begin{abstract}
A new integration technique is presented for systems of linear
partial differential equations (PDEs) for which syzygies can be
formulated that obey conservation laws. These syzygies come for free
as a by-product of the differential Gr\"{o}bner Basis computation.
Compared with the more obvious way of integrating a single equation
and substituting the result in other equations the new technique
integrates more than one equation at once and therefore introduces
temporarily fewer new functions of integration that in addition
depend on fewer variables. Especially for high order PDE systems in
many variables the conventional integration technique may lead to an
explosion of the number of functions of integration which is avoided
with the new method. A further benefit is that redundant free
functions in the solution are either prevented or that their number
is at least reduced.
\end{abstract}

\section{A critical look at conventional integration} 
In this paper a new integration method is introduced that 
is suitable for the computerized solution of systems of linear PDEs
that admit syzygies. In the text we will call the integration of
single exact differential equations, i.e.\ equations which are total
derivatives, the `conventional' integration method (discussed, for
example, in \cite{Wol99e}). To highlight the difference with the new
syzygy based integration method we have a closer look at the
conventional method first. About notation: To distinguish symbolic 
subscripts from partial derivatives we indicate partial derivatives
with a comma, for example, $\partial_{xy} e_i = e_i,_{xy}$.

To solve, for example, the system 
\begin{eqnarray}
0 & = & f,_{xx}        \label{ie1} \\
0 & = & xf,_y + f,_z   \label{ie2}
\end{eqnarray}
for $f(x,y,z)$ one would, at first, integrate (\ref{ie1})
with 2 new functions of integration $g(y,z), h(y,z)$, 
then substitute 
\begin{equation}
f = xg + h             \label{ie3}
\end{equation}
into (\ref{ie2}), do a separation with
respect to different powers of $x$ to obtain the system
\begin{eqnarray*}
0 & = & g,_y          \\ 
0 & = & g,_z + h,_y   \\ 
0 & = & h,_z             
\end{eqnarray*}
and solve that to get the solution
\[f=x(az+b) - ay + c, \ \ \ \ \ \ \ \ \ a,b,c=\mbox{const}. \]
The main gain of information on which the overall success was based 
did happen after the substitution
at the stage of separating (\ref{ie2}) into 3 equations.
The integration of (\ref{ie1}) itself did not provide new information.
The equation $0=f,_{xx}$ is more compact than $f = xg + h$
and equally well usable in an ongoing elimination process (Gr\"{o}bner
Basis computation). (Similarly, in this sense,
$f(x)=a\sin(x)+b\cos(x)$ would not provide
new information compared to $0=f''+f$ as $\sin$
and $\cos$ are only defined as solutions of this ODE.)
The main conclusion is:
{\it The integration of a single equation does not necessarily imply
progress in the solution of a system of PDEs, especially if a direct
separation does not become possible as the result of substituting
a computed function.}

This is the case in the example
\begin{eqnarray}
0 & = & f,_{yzz}   \;\;\;\;\;\;\;\;\;\;\;\, ( =: e_1 )  \label{e1} \\
0 & = & f,_{xx} + f,_z .       \;\;\;\;     ( =: e_2 )  \label{e2} 
\end{eqnarray}
discussed in more detail in the next section. Integration of (\ref{e1}) to 
$f=g_1(x,y) + z g_2(x,y) + g_3(x,z)$ and substitution into (\ref{e2}) 
does not yield a separable equation and is therefore not as straight forward
to utilize as in the first example.

There is another problem with the conventional method which seems
insignificant at first sight but becomes severe for high order PDE systems
in many independent variables, for example in the application
in section \ref{reallife}.

Substituting $f=g_1(x,y)+zg_2(x,y)+g_3(x,z)$ into (\ref{e2}) as done in section
(\ref{conven}) and finding the general solution for $g_1, g_2, g_3$
is, strictly speaking, a different problem from
finding the general solution for $f$
of (\ref{e1}), (\ref{e2})! The general solution for $g_1, g_2$,
as determined in section \ref{conven}, 
will involve among other functions the two essential
free functions $g_6(x), g_7(x)$. From the point of view of the
original system (\ref{e1}), (\ref{e2}) these are redundant 
functions as they can be absorbed by $g_3$. Redundancy is an inherent
problem of the conventional integration method which has nothing to do
of how efficient the remaining system after integration and
substitution is solved. In section \ref{redund} this issue is
discussed in more detail. 

With the new syzygy based integration the situation is very different.
Here the decision whether to integrate is based on syzygies, i.e.\
on relations {\it between} equations, like
\[0 = (\partial_x^2 + \partial_ z) e_1 -
       \partial_y\partial_z^2 e_2 \]
in the last example and is not based on the form of a {\it single} equation.
This extra information content coming from the syzygies allows the method
to perform useful integrations for systems like
(\ref{e1}), (\ref{e2}) with an instantly useful result. 
As will be explained further below,
syzygy based integration does not only integrate one single equation
at a time, but in a sense, it performs an integration which is
compatible with all the equations involved in the syzygy.
(More exactly, it integrates all equations $0=P^i$ at once one time where
$P^i$ are the components of the conserved current of the conservation
law of the syzygies.)

This restrictive 'compatibility constraint' has the effect that
the integral involves fewer new functions
of integration which furthermore depend on fewer variables.
Consequently fewer new functions have to be computed later on
which shortens the computation. Also, fewer redundant functions
are generated which not only avoids the explosion of the number of 
intermediately generated functions but also simplifies the final 
solution. These effects are especially important for high order PDEs
in many variables as explained in section \ref{redund}.

The above distinctions between both integration techniques are
not purely academic. Section \ref{threeruns} describes how 
integrations can be combined with eliminations.
To apply integrations early in the solution process is not new.
This strategy has been pursued by
the program {\sc Crack} for nearly 2 decades.
What is interesting and new is how much more beneficial the syzygy based
integration proves to be compared with conventional integration. 
In section \ref{threeruns} such a comparison has been made. One
problem has been solved 3 times with a combination of different
modules, including elimination and conventional and syzygy based
integration. The 3 runs differ only in the priority of applying these
modules and were compared by their running times as well as the number
of redundant functions in the final solution.\vspace{4pt}  

{\it About the remainder of the paper}  \vspace{4pt} \\
In section \ref{general} the algorithm is
described in general and an overview is provided. \vspace{4pt} \\
Using the
information content of syzygies in the form of conservation laws 
seems to be the most direct and useful way but it is not the only one possible.
In section \ref{curl} a variation of the algorithm is explained which
is based on vanishing curls of syzygies. \vspace{4pt} \\
Different aspects of the computation of conservation laws
for syzygies are the subject of the following section. \vspace{4pt} \\
The redundancy problem mentioned above is looked at in detail in
section \ref{redund}. \vspace{4pt} \\
Even though
conservation laws of syzygies might be known, it may not be
advantageous to use them if the aim is the exact solution of the original
PDE-system. In section \ref{useless} examples are given. \vspace{4pt} \\
A short description of how syzygies are recorded in section 8 is followed by
section \ref{reallife} introducing
the `real-life' application which led to the development of
syzygy based integration.
In three computer runs it is shown that this integration method and
elimination can be naturally combined for the solution of linear PDE
systems. \vspace{4pt} \\
In the following section the introductory example is 
continued and both integration methods are compared.

\section{An introductory example} \label{exmpl} 
We continue the above example to explain the basic mechanism of syzygy
based integration. A more complex example is given in section
\ref{reallife}. 

\subsection{Treated with the new method} \label{intro1}
In applying integrability conditions for PDEs systematically, i.e.\
in computing a differential Gr\"{o}bner basis, identities between
equations $0=e_a$ will result that take the form of differential
expressions with the $e_a$ as dependent variables.

We consider the simple system (\ref{e1}), (\ref{e2}), i.e.\
\begin{eqnarray*}
0 & = & f,_{yzz}   \;\;\;\;\;\;\;\;\;\;\;\, ( = e_1 ) \\
0 & = & f,_{xx} + f,_z .       \;\;\;\;     ( = e_2 ) 
\end{eqnarray*}
Assuming, for example, a total ordering $>_o$ of derivatives that implies
$\partial_x >_o \partial_z$ and $\partial_y >_o \partial_z$,
a differential Gr\"{o}bner Basis computation would first
eliminate $f,_{xxyzz}$ through cross-differentiation:
\begin{equation} 0 = e_2,_{yzz} - e_1,_{xx} = f,_{yzzz} \;\;\;\;\;\;( =: e_3 )
\label{e3a} \end{equation}
then a substitution of $f,_{yzz}$ using $e_1$ yields
\[ 0 = e_3 - e_1,_z  \]
and a substitution of $e_3$ using (\ref{e3a})
provides the identity
\begin{equation}
 0 = e_2,_{yzz} - (e_1,_{xx} + e_1,_z).             \label{e3}
\end{equation} 
The choice of ordering does not matter here. Any ordering would have
resulted in identity (\ref{e3}).

In this paper we concentrate ourselves to the integration of
syzygies, like (\ref{e3}), which either have the form of a 
divergence or can be combined linearly to give a divergence
$0 = D_i P^i$ with suitable vector components $P^i(e_k)$ that are
differential expressions in the $e_k$. Only in section \ref{curl} we
outline a variation of this principle to deal with a vanishing
curl of syzygies.  

The computation of conservation laws of syzygies has
several aspects: how to do it in general, why the computation of
conservation laws for syzygies is a relatively simple task and how to
do it in less generality but much faster. In the interest of a compact
example we postpone this discussion to section \ref{conlaw}.

There are different ways to write (\ref{e3}) as a divergence.
We choose any one with as few as possible components (here two: $P^x, P^z$).
This preference is justified towards the end of this section
below equation (\ref{id9c2}). The question how conservation laws
with fewer components are computed is described in section \ref{conlaw} as well.

We obtain:
\begin{eqnarray}
0 & = & - e_1,_{xx} + \, (e_2,_{yz} - e_1),_z  \label{id1}   \\
  & = & P^x,_x + P^z,_z                     \label{id2}   \\
  & = & (-f,_{xyzz}),_x + \,(f,_{xxyz}),_z.     \label{id3}  
\end{eqnarray} 
In the following we will use the
vector $P^i$ in two representations, first in terms of $e_i$, in our example
from the syzygy (\ref{id1}):
\begin{equation}
P^x = - e_1,_x,  \;\;\;\;P^z = e_2,_{yz} - e_1  \label{P1}
\end{equation}
and second the representation of $P^i$ in terms of the 
function $f$, in our example from the identity (\ref{id3}):
\begin{equation}
P^x = -f,_{xyzz},  \;\;\;\;P^z = f,_{xxyz}.      \label{P2}
\end{equation}
With $P$ satisfying the conservation law condition (\ref{id2}) 
we can write $P$ as a 2-dim.\ curl  
\begin{equation}
 P^x = - Q,_z,  \;\;\;\; P^z = Q,_x        \label{id5} 
\end{equation}
for some potential $Q$. Using for $P^i$ the representation (\ref{P2})
we identify
\[Q = f,_{xyz}. \]                
The existence of differential expressions in unknowns, say $f^\alpha$,
for the potential
$Q$ is guaranteed because all syzygies and all their consequences
like $0 = D_i P^i$ are satisfied identically for any $f^\alpha$.
In the appendix B an algorithm {\sc DivInt} is given that computes potentials
$Q^{ij}(f^\alpha)$ in general for an arbitrary number of independent
variables.

To do the next step in this example,
we are reminded that expressions $P^i(e_j)$ are linear homogeneous in
the $e_j$ and that they therefore must be zero, i.e.\ $P^x=P^z=Q,_x=Q,_z=0$. 
This means that
$Q$ is independent of $x,z$, giving $Q=c_1(y)$ and the new equation
\begin{equation}
0 = Q - c_1 = f,_{xyz} - c_1 \;\;\;\;\;\; ( =: e_4 )   \label{e4}
\end{equation} 
with the new function of integration $c_1=c_1(y)$. 

Apart from the integral (\ref{e4}) we also get new syzygies.
Having on one hand expressions for $P^i$ in terms of $e_1, e_2$ due to
equations (\ref{P1}) and on the other hand $P^i$ in
terms of $Q,_j$ from equations (\ref{id5})
and $Q$ in terms of $e_4$ from equation   
(\ref{e4}) we get two new identities
\begin{eqnarray}
0 & = & P^x + Q,_z = - e_1,_x + e_4,_z           \label{id7} \\
0 & = & P^z - Q,_x = e_2,_{yz} - e_1 - e_4,_x.   \label{id8}
\end{eqnarray}
As equation $e_1$ turns up algebraically in at least one of the new
identities, this equation $0=e_1$ is redundant and can be dropped. 
Redundancy of an original equation due to integration 
need not always be the case but it is the case in this
example because at least one of
$P^x$ and $P^z$ happens to be algebraic in $e_1$ (in this case $P^z$).
Identity (\ref{id8}) has already conservation law form.
Substituting $e_1$ from identity (\ref{id8}) into (\ref{id7})
preserves this form:
\begin{equation}
0 = (- e_4,_x),_x \, + \, (e_2,_{xy} - e_4),_z. \label{ide8a}
\end{equation}
This completes one syzygy based integration step. Because the new
system of equations $0=e_2=e_4$ obeys the syzygy (\ref{ide8a})
which has a conservation law form with only 2 components $P^x, P^z$
we can start another integration step {\em without having to do
a differential reduction or cross differentiation step}. 
It turns out there are in total 3 more very similar syzygy integration steps
to be performed which are summarized in appendix A. After these 3
steps the remaining system to solve consists of the 2 equations
\begin{eqnarray}
0 & = & f,_{xx} + f,_z  \;\;\;\;\;\;\;\;\;\;\;\;\;\;\;\;\;
\;\;\;\;\;\;\;\;\;\;\;\;\;\;\;\;\;\;\;\;\;\;
\;\;\;\;\;\;\;\;\;\;\;\;\;\;\;\; ( = e_2 ) \label{e2a} \\
0 & = & f,_y + \frac{x^3}{6} c_1 - \frac{x^2}{2}c_2 - xzc_1 + zc_2 - xc_3 - c_4
\;\;\;\;\;\; ( = e_7 )  \label{e7a}
\end{eqnarray}
which satisfy the identity
\begin{equation}
   0 = - e_2,_y + e_7,_{xx} + e_7,_z.               \label{id9c2}
\end{equation} 
This is a divergence too but now in three differentiation variables. 
With three non-vanishing $P^i$ the condition $0=D_i P^i$ has the
solution $P^i = D_j Q^{ij}$ with more than one non-vanishing $Q^{ij}$ 
and the condition $0=P^i=D_j Q^{ij}$ has the solution
$Q^{ij} = R^{ijk},_k$
with free functions $R^{ijk}(x^n)=R^{[ijk]}(x^n)$
where $^{[ijk]}$ stands for total antisymmetrization.
In three dimensions this introduces one new function 
$R(x^n)=R^{xyz}$ through 
$Q^{xy}=R,_z, \;\; Q^{yz}=R,_x,$ and $Q^{zx}=R,_y$.
By performing a syzygy based integration again we would solve the remaining
equations (\ref{e2a}),(\ref{e7a}) for one function
$f$ but also introduce one new unknown function $R$ of all variables and
therefore not make real progress. This is demonstrated in the first
example in section \ref{useless}.
These considerations explain why we try to find conservation
laws of syzygies with as few as possible non-zero $P^i$.

We return to our example and decide to integrate $0=e_7$ (i.e.\
(\ref{e7a})) conventionally because
\begin{itemize}
\item identity (\ref{id9c2}) can not be written as a divergence with only 2
  terms and
\item equation (\ref{e7a}) can be integrated conventionally with respect to
  {\it only one} integration variable, so we will not introduce redundant
  functions as discussed in the introduction and in section \ref{redund}.
\end{itemize}
To $y$-integrate equation (\ref{e7a}) we
introduce four new functions $d_1(y),\ldots, d_4(y)$ through $c_i=d_i,_y$ and
one new function $d_5=d_5(x,z)$ and obtain
\begin{equation}
 f = - \frac{x^3}{6} d_1 + \frac{x^2}{2}d_2 + xzd_1 - zd_2 + xd_3 + d_4 + d_5
     \label{e8}
\end{equation} 
with the only remaining equation (\ref{e2a}) now taking the shape
\begin{equation}
0 = d_5,_{xx} + d_5,_z.  \label{e9}
\end{equation} 
A single equation does not have syzygies and the method can not be
applied further. What we achieved is the integration of equation
(\ref{e1}) and the change of equation (\ref{e2}) for 3 independent
variables into equation (\ref{e9}) for 2 variables.

\subsection{The same example in a conventional treatment} \label{conven} 
For comparison, we solve the system (\ref{e1}), (\ref{e2}) again, this time in
the conventional direct way. After integrating (\ref{e1}) to
\begin{equation}
f = g_1(x,y) + z g_2(x,y) + g_3(x,z)  \label{e10}
\end{equation} 
and substitution of $f$ the equation (\ref{e2}) reads
\begin{equation}
0 = g_1(x,y),_{xx} + z g_2(x,y),_{xx} + g_3(x,z),_{xx} + 
    g_2(x,y) + g_3(x,z),_z.  \label{e11}
\end{equation} 
In equation (\ref{e11}) there is no function that
does depend on all variables and each
variable does occur in at least one function. An algorithm for such 
`indirectly separable equations' (ISEs) is contained in the package 
{\sc Crack} (see \cite{HPR} and sub-section \ref{threeruns}).
These equations undergo a series of differentiations and
divisions (producing a list of divisors) 
\begin{itemize}
\item to eliminate all functions of some variable, 
\item to do a direct separation with respect to this variable, and
\item to use the same list of divisors now in reverse order
      as integrating factors to back-integrate the equations which
      resulted from direct separation. 
\end{itemize}
In the case of equation (\ref{e11}) a single $y$-differentiation 
eliminates $g_3$ and allows a direct $z$ separation (as $g_1, g_2$ are
independent of $z$) giving 
$0= g_2(x,y),_{xxy},\; 0=g_1(x,y),_{xxy} + g_2(x,y),_y$ and through
back-integration with respect to $y$ further
\begin{eqnarray}
0 & = & g_2(x,y),_{xx} + g_4(x)                           \label{e12} \\
0 & = & g_1(x,y),_{xx} + g_2(x,y) + g_5(x)             \label{e13} \\
0 & = & g_3(x,z),_{xx} + g_3(x,z),_z - z g_4(x) - g_5(x)  \label{e14}
\end{eqnarray}
with new functions of integration $g_4, g_5$. Renaming
$g_4=g_6(x),_{xxxx},\; g_5=g_7(x),_{xx}$
and integrating equations (\ref{e12}), (\ref{e13}) gives
\begin{eqnarray}
g_2 & = & - g_6(x),_{xx} - xg_8(y) - g_9(y)    \label{e15a} \\
g_1 & = &   g_6(x) + \frac{x^3}{6}g_8(y) + \frac{x^2}{2}g_9(y) 
          + xg_{10}(y) + g_{11}(y) - g_7(x) \label{e16a} \\
0 & = &   g_3(x,z),_{xx} + g_3(x,z),_z 
          - z g_6(x),_{xxxx} - g_7(x),_{xx}  \label{e17a} \\
f & = &   g_3(x,z) + g_6(x) + \frac{x^3}{6}g_8(y) 
          + \frac{x^2}{2}g_9(y) + xg_{10}(y) + g_{11}(y) - g_7(x)
          \nonumber \\
  &   &   - z(g_6(x),_{xx} + xg_8(y) + g_9(y)) .  \label{e18a}
\end{eqnarray}
The solution (\ref{e18a}) is identical to (\ref{e8}) and the remaining
condition (\ref{e17a}) is identical to (\ref{e9}) if we drop the
redundant functions $g_6, g_7$ which can be absorbed by $g_3$
and substitute $g_8=-d_1, g_9=d_2, g_{10}=d_3, g_{11}=d_4, g_3=d_5$.
A method to recognize redundancy is described in \cite{WBM}. It
involves the solution of an over-determined system of equations
which involves even more effort.

The introduction of redundant functions $g_6, g_7$ in the conventional
method was unavoidable because after reaching system (\ref{e12}) -
(\ref{e14}) with the task to compute $g_1,\ldots, g_5$ 
the information was lost that, strictly speaking, not the most general
expressions for $g_1,\ldots, g_5$ need to be computed but only the most
general expression for $f=g_1(x,y) + z g_2(x,y) + g_3(x,z)$. 
Setting $g_6=g_7=0$ would be a restriction for $g_2$ and $g_1$ in
(\ref{e15a}), (\ref{e16a}) but is not a restriction for $f$ in (\ref{e18a}).

\section{The algorithm in general} \label{general}  
In our notation $x^i, \;\,i=1,\ldots,p$ are the
independent variables and $f^\alpha$ are the unknown functions
which do not need to depend on all $x^i$.
These functions satisfy equations $0=e_a(x^n,f^\alpha_J)$ 
where $_J$ is a multi-index (standing, for example, for $_{112}$, i.e.\
$\partial^2_{x^1}\partial_{x^2}$) and where $f^\alpha_J$ 
stands for a possible dependence on $f^\alpha$ and any partial
derivatives of $f^\alpha$. Total derivatives
appear as $D_i$. Summation is performed over identical indices.

The following description is summarized in the overview underneath.
The number(s) at the start of each item refer to the line number
of the corresponding step in the overview.
\begin{description}
\item[(\ref{over1}),(\ref{over2}):] 
For a given system of differential equations (\ref{over1}) the 
investigation of integrability conditions (e.g.\ Gr\"{o}bner 
basis computation) yields identities 
(\ref{over2}), called  syzygies. In these syzygies the $e_k$ take
the role of dependent variables.
The program {\sc Crack} has been used to compute syzygies 
for examples presented in this paper but many other computer
algebra programs are available (for example, RIF \cite{Reid2},
diffalg \cite{BLOP1},\cite{Hub1},\cite{Hub2},
diffgrob2 \cite{Mans}) although only few generate syzygies
automatically.
\item[(\ref{over3}):] 
To find conservation laws of syzygies
one either can perform a
more expensive but general search by using the package 
{\sc ConLaw} \cite{Wol99a} or other computer algebra
software, or one can do a more specialized, less general but faster
computation as described in section \ref{faster}.
In the conservation laws as in the syzygies
the dependent variables are the $e_k$.

In order to introduce as few as possible new functions through a
syzygy based integration, one aims at conservation laws with as few as
possible non-zero $P^i$ (see discussion towards the end of section
\ref{intro1}). Possible methods to achieve this are described in
section \ref{choosing}.

Most often syzygies are very simple expressions and already have a
conservation law form. Computing conservation laws is not fully 
algorithmic but it is argued in section \ref{underdet} that this task 
is relatively simple for under-determined systems of syzygies.
\item[(\ref{over3a}):] 
If a conservation law for the syzygies is known then the following
steps can definitely be performed. The question is only whether it is
beneficial for the purpose of the computation.
If one has found a conservation law with only 2 components $P^i$
then the integration will introduce just one new constant and will
always be beneficial. If the conservation law has 3 or more
components $P^i$ then at least one new function of all
variables will be introduced. In that case,
if the purpose of the integration is the
solution of the PDE system (\ref{over1}) then
one would have to balance how many functions one can solve for
due to the new integrated equations (\ref{over6}) against how many new
functions are introduced and possible decide not to continue.
Examples for syzygy based integrations which are useful from the point of solving PDE-systems 
and others that are not are shown in section \ref{useless}.
If usefulness can not be decided at this stage then the integration should be
performed and decided afterwards. The computational complexity of the
integration, i.e.\ of the algorithm {\sc DivInt} is very low.
\item[(\ref{over4}):] 
In the computed conserved currents $P^i(x,e_k)$ we replace
the equation names $e_k$ by their expressions (\ref{over1}) in terms of
$x, f^\alpha$. 
\item[(\ref{over5}):] 
The resulting $P^i(x,f^\alpha)$ in (\ref{over4}) 
are the input to the algorithm {\sc DivInt} (given in the appendix B) to
compute a special solution for the potentials 
$Q^{ij}=Q^{[ij]}(x,f^\alpha)$ satisfying $P^i = D_j Q^{ij}$. 
Here again $^{[ij]}$ stands for anti-symmetrization.
{\sc DivInt} works because the kernel of a divergence
$D_i P^i$ is a curl $D_jQ^{ij}$ with $Q^{ij}=-Q^{ji}$
and because $0=D_i P^i$ is satisfied identically in all $f^\alpha$ and
their derivatives.
\item[(\ref{over5a}):] 
Because the syzygies $0=\Omega_m(x,e_k)$
are linear homogeneous expressions in the
$e_k$, therefore $D_i P^i$ being a linear homogeneous expression in the $\Omega_m$
is also a linear homogeneous expression in the
$e_k$. Hence the $P^i$ are linear homogeneous expressions in the
$e_a$. Consequently, we have $0 = P^i$ in the space of solutions of the
original equations.\footnote{When computing a differential Gr\"{o}bner
Basis the equations in the final basis are also only differential
consequences of the initial equations and one would not want to delete
them. Here the situation is different. $0=e_m$ has been integrated and
can be deleted if $e_m$ occurs algebraically in other syzygies.}
\item[(\ref{over6}):] 
On the other hand, the algorithm {\sc DivInt} 
computes expressions $Q^{ij}$ satisfying $P^i = D_j Q^{ij}$ identically
and therefore $0=D_j Q^{ij}$ in the space of solutions of the original
equations. The general solution of this condition for the $Q^{ij}$ 
is shown in (\ref{over6}) and is the result of the whole computation.
Its form depends on the number $p$ of non-vanishing components $P^i$:
for $p=2$ a single constant of integration $R$ is introduced
for $p>2$ one or more functions $R^{ijk}(x)$ are introduced.
\item[(\ref{over7}):] 
The formal integration of $0 = D_j Q^{ij}$ gives new 
equations whose right hand sides are abbreviated by $e_{ij}$.
\item[(\ref{over8}):] 
We are instantly able to formulate syzygies which
these new equations $0=e_{ij}$ satisfy.
\item[(\ref{over9}),(\ref{over9a}):] 
If any one of them can be solved for one $e_m$ 
(as indicated in (\ref{over9})) then $e_m=\omega$
can be substituted in other syzygies and the original equation
$0=e_m(x,f^\alpha)$ can be deleted as it is a consequence
of the equations $e_k,e_{ij}$ in $\omega(x,e_k,e_{ij},_j)$.
\item[(\ref{over10}):] 
1. As new syzygies have been generated in (\ref{over8}) there is a chance
that anyone of them has already a conservation law form, like (\ref{id7}).\\
2. The substitution of a redundant equation in step (\ref{over9}) may
also lead to a syzygy in conservation law form, either in the other
newly generated syzygy or in any other syzygies. \\
3. Finally, there is always the possibility that the new syzygies combined with
other syzygies take a conservation law form. This would have to be
found out by a computation, for example using the program {\sc ConLaw}.\vspace{-12pt}
\end{description}
\begin{eqnarray}
& \mbox{Given system:}\ \ \ \ \ \ \ \ \ \ & 0 = e_k(x,f^\alpha) 
\label{over1} \vspace{4pt}\\
\!\!\!\!\!\!\!\!\!\!\mbox{{\sc Crack}}\ \ \, \rightarrow &
\mbox{Syzygies:}\ \ \ \ \ \ \ \ \ \ \ \ \ \ \ \  & 0 = \Omega_m(x,e_k)
\label{over2} \vspace{4pt}\\
\!\!\!\!\!\!\!\!\!\!\mbox{{\sc ConLaw}} \rightarrow 
&\mbox{Cons. law form:}\ \ \ \ \ \ \  & 0 = D_i P^i(x,e_k), 
\label{over3} \vspace{4pt}\\
&\mbox{Is CL useful?} \ \ \ \ \ \ \ \ \ \ & \mbox{If not then stop.}
\label{over3a} \vspace{4pt}\\
&\mbox{Conserved current:}\ \ \ & 
P^i=P^i(x,e_k)|_{e_k \rightarrow e_k(x,f^\alpha)}=P^i(x,f^\alpha) 
\label{over4} \vspace{4pt}\\
\!\!\!\!\!\!\!\!\!\!\mbox{{\sc DivInt}}\ \  \rightarrow 
&\mbox{New potentials:}\ \ \ \ \ \ \ \, & P^i(x,f^\alpha) = D_j Q^{ij} \ \mbox{ with } \
                            Q^{ij}=Q^{[ij]}(x,f^\alpha) \label{over5} \vspace{4pt}\\
&\mbox{Integration of:}\ \ \ \ \ \ \ \ \, & 0 = P^i = D_j Q^{ij} 
                            \label{over5a}  \vspace{4pt}\\
&\mbox{to new integral(s):}\ \ \ \ &
Q^{ij}(x,f^\alpha) \!=\! \left\{ 
           \begin{array}{lll} 
               \!\!\!R=\mbox{const} & \!\!\!\mbox{in \,\ \ 2 dim} \vspace{4pt}\\
               \!\!\!R^{ijk},_k \mbox{with}\;R^{ijk}\!=\!R^{[ijk]}(x)
                            & \!\!\!\mbox{in $>$2 dim} 
           \end{array} 
         \right.   \label{over6} \vspace{4pt}\\
&\mbox{New equation names:} &  
0 = \left\{  \begin{array}{l} Q^{ij}(x,f^\alpha)-R \\ 
                              Q^{ij}(x,f^\alpha)-R^{ijk},_k
\end{array} \right\} =:e_{ij} \label{over7} \vspace{4pt}\\
&\mbox{New syzygies:}\ \ \ \ \ \ \ \ \ \  & \rightarrow 0 = P^i(x,e_k) - e_{ij},_j 
                     \label{over8} \vspace{4pt}\\
&\mbox{Redundancies?}\ \ \ \ \ \ \ \ \, & e_m=\omega(x,e_k,e_{ij},_j) \rightarrow 
                               \label{over9} \\
&                          & - \;\mbox{substitution of}\;e_m=\omega \; 
                            \mbox{in any syzygy}  \label{over9a} \\
&                          & - \;\mbox{deleting equation}\; 0=e_m \nonumber \\
& \mbox{return to the} \ \ \ \ \ \ \ \ \ \  & 
\mbox{determination of conservation laws for syzygies}  \label{over10}
\end{eqnarray} 
The continuation of the introductory example in appendix A is itemized
similar to the description above. This allows the reader to go through
an example and compare it with the overview step by step.

\section{An integration based on curls of syzygies} \label{curl} 
The described ansatz of extracting information out of syzygies in order
to do integrations is not the only possible way. 
In this section we want to provide a different integration method,
this time based on vanishing curls of syzygies. 
We will see that it is even more
effective than divergence based integration but the required structure
of the system of syzygies is more special which is the reason why it has
not been implemented in {\sc Crack}. Also, the computation of
conservation laws for syzygies was implemented so far
because computer
programs, like {\sc ConLaw}, are available to compute conservation laws
and because the existence of conservation laws is a
relative weak condition for syzygies. The method based on curls is
shown in the following overview.

\[ \!\!\begin{array}{ll}
\mbox{Given system:}      & 0 = e_k(x,f^\alpha) \vspace{4pt}\\
\mbox{Syzygies:}          & 0 = \Omega_m(x,e_k) \vspace{4pt}\\
\mbox{Vanishing curl cond.:} & 0 = D_j P^{ij} \ 
     \mbox{ with } \ P^{ij}=P^{[ij]}(x,e_k),\vspace{4pt}\\
\mbox{Curl free tensor:}  & 
P^{ij}=P^{ij}(x,e_k)|_{e_k \rightarrow e_k(x,f^\alpha)}=P^{ij}(x,f^\alpha) \vspace{4pt}\\
\mbox{New potentials:}    & P^{ij}(x,f^\alpha) = D_k Q^{ijk} \ \mbox{ with } \
                            Q^{ijk}=Q^{[ijk]}(x,f^\alpha) \vspace{4pt}\\
\mbox{Integration of:}    & 0 = P^{ij} = D_k Q^{ijk} \vspace{4pt}\\
\mbox{to new integral(s):}&
Q^{ijk}(x,f^\alpha) \!=\! \left\{ 
           \begin{array}{lll} 
               \!\!\!R=\,\mbox{const} & \!\!\!\mbox{in \ \ 3 dim} \vspace{4pt}\\
               \!\!\!\!\!\begin{array}{l} R^{ijkl},_l \;\mbox{with}\; 
                                       R^{ijkl}\!=\!R^{[ijkl]}(x) \end{array}\!
                            & \!\!\!\mbox{in $>$3 dim} 
           \end{array} 
         \right.    \vspace{4pt}\\
\mbox{New equation names:}  &  
0 = \left\{  \begin{array}{l} Q^{ijk}(x,f^\alpha)-R \\ 
                              Q^{ijk}(x,f^\alpha)-R^{ijkl},_l
\end{array} \right\} =:e_{ijk} \vspace{4pt}\\
\mbox{New syzygies:}      & \rightarrow 0 = P^{ij}(x,e_a) - e_{ijk},_k \vspace{4pt}\\
\mbox{Redundancies?}      & e_m=\omega(x,e_k,e_{ijk},_k) \rightarrow
                            \;\mbox{substitution of}\;e_m       \\
                         & - \;\mbox{substitution of}\;e_m=\omega \; 
                            \mbox{in any syzygy} \\
                         & - \;\mbox{deleting equation}\; 0=e_m \\
\mbox{return to the }  & \mbox{determination of vanishing curls or
                               divergences for syzygies}
\end{array} \]

The superficial difference between divergence and curl based integration
is that $P,Q,R$ have one extra index for the curl based method.
This method also needs at least 3 independent variables.
The following two examples involve each 4 independent variables and
allow a closer comparison of both methods.

{\em A typical example:}\\ 
For 4 unknown functions $a,b,c,d$ depending on $x,y,z,t$ a system of 6 equations
\[ \begin{array}{rcclrcclrccl}
0&=&d,_z-c,_t&(=:e_{xy})\,,\;\;&\;\; 0&=&b,_t-d,_y&(=:e_{xz})\,,\;\;&\;\; 0&=&c,_y-b,_z&(=:e_{xt}) \\
0&=&d,_x-a,_t&(=:e_{yz})\,,\;\;&\;\; 0&=&a,_z-c,_x&(=:e_{yt})\,,\;\;&\;\; 0&=&b,_x-a,_y&(=:e_{zt})
\end{array} \]
is given. It has syzygies
\[\begin{array}{lcr}
0&=& e_{xy,y}+e_{xz,z}+e_{xt,t} \\
0&=&-e_{xy,x}+e_{yz,z}+e_{yt,t} \\
0&=&-e_{xz,x}-e_{yz,y}+e_{zt,t} \\
0&=&-e_{xt,x}-e_{yt,y}-e_{zt,z} 
\end{array}  \]
which take the form of a vanishing curl: $0=D_jP^{ij}$ for
$P^{ij}=e_{ij}$ leading to potentials $Q^{ijk}$
\[Q^{xyz}=d,\;\;\;Q^{txy}=c,\;\;\;Q^{xzt}=b,\;\;\;Q^{ytz}=a \]
and a single new free function of integration $R^{xyzt}=g(x,y,z,t)$.
The resulting integrals are
\[ a=g,_x\,,\;\;\;\;b=g,_y\,,\;\;\;\;c=g,_z\,,\;\;\;\;d=g,_t .\]

{\em A related example for a conservation law syzygy:}\\ 
In comparison, the typical example using a conservation law syzygy in
4 independent variables would involve 6 unknown functions
$a,b,c,d,f,g$ and 4 equations, so a less over-determined system:
\[ \begin{array}{rcclrcclrccl} 
0&=&\;\;a,_y+b,_z+c,_t&(=:e_1)\,,\;\;&\;\;0&=&-a,_x+d,_z+f,_t&(=:e_2)\\
0&=&-b,_x-d,_y+g,_t&(=:e_3)\,,\;\;&\;\;0&=&-c,_x-f,_y-g,_z&(=:e_4).
\end{array} \]
The conservation law $0=e_1,_x+e_2,_y+e_3,_z+e_4,_t$ gives $P^i=e_i$
and potentials 
\[Q^{xy}=a,\;\;\;Q^{xz}=b,\;\;\;Q^{xt}=c,\;\;\;Q^{yz}=d,\;\;\;Q^{yt}=f,\;\;\;Q^{zt}=g. \]
The resulting integrals are
\[ a=r,_z-s,_t\,,\;\;\;\;b=u,_t-r,_y\,,\;\;\;\;c=s,_y-u,_z\,,\;\;\;\;
   d=r,_x-w,_t\,,\;\;\;\;f=w,_z-s,_x\,,\;\;\;\;d=u,_x-w,_y \]
with new arbitrary functions $r,s,u,w$.

If both methods would be applicable, i.e.\
if the system of syzygies would provide a vanishing divergence and a vanishing
curl then one would prefer the curl based integration because it makes
use of more syzygies.

The last two examples look very artificial but one could exchange the
unknown functions $a,b,c,\ldots$ by any functionally independent
expressions, each of them involving at least one different function,
and the computations and results would be unchanged.

The remainder of the paper is concerned with divergence based integration.

\section{How to find conservation laws of syzygies} 
\label{conlaw}
In order to find a combination of syzygies that is a divergence one
could apply computer algebra programs {\sc ConLaw} as described in
\cite{Wol99a}, \cite{WBM} by regarding the syzygies as the
equations and the $e_a$ as unknown functions. In the following
subsections we discuss why computing conservation laws of syzygies is
simpler than computing conservation laws in general, how one can find
conservation laws with fewer components than independent variables and
how conservation laws for syzygies are determined in {\sc Crack}.

\subsection{Under-determination of syzygies} \label{underdet}
If one interprets syzygies as PDEs for unknowns $e_k$, then the
original equations $e_k=e_k(x^i,f^{\alpha}_J)$ are special solutions of
these syzygies where the $f^\alpha$ play the role of arbitrary
functions in these solutions. Because at least one of the $f^\alpha$
depends on all variables $x^i$ (otherwise the original system consists
only of ISEs to be treated differently, not by checking integrability
conditions), the syzygies must be an under-determined 
PDE-system for the unknowns $e_k$. Computing
conservation laws for under-determined systems of PDEs is an even more
over-determined problem. The conservation law conditions have to be
satisfied identically in a jet space with coordinates $x^n, e_a$ and
all partial derivatives of all $e_a$. 
The more $e_a$ occur in the syzygies the more
restrictive are their conservation law conditions. 
Another way to see this is that 
conditions for integrating factors to give conservation laws are
obtained by applying the variational derivative (Euler-Lagrange
operator) to the product of integrating factors and syzygies (see
\cite{Olver1986}). Because there is one Euler operator for each $e_a$
we get as many conditions as there are different $e_a$.
Finally, the more over-determined a system of
conditions is, the easier it is to solve. Therefore the subtask of
computing conservation laws of systems of syzygies is usually not a
problem. 

\subsection{Choosing between different 
syzygy conservation laws} \label{choosing} 
The integration of a syzygy $0 = D_i P^i$ with two derivatives
$0=D_x P^x + D_y P^y$ is always
useful but not necessarily the integration of a syzygy with
more than 2 derivatives because there is at least one new function
of integration of all variables (see the example in section \ref{useless}). 
Sometimes there is a choice
allowing to write a syzygy in different forms, for example 
\[0=e_1,_x + (e_2,_x)_y + e_3,_z\] can also be written as 
\[0=(e_1 + e_2,_y),_x + e_3,_z.\] 
To find out whether a conservation law with
fewer derivatives exists one has two options.
First, one can make an ansatz for the conservation law with fewer
derivatives and solves the resulting conditions (for example, with
the programs {\sc ConLaw1} or {\sc ConLaw3}). 
Alternatively, one computes the most general conservation law 
involving arbitrary functions.
If a conservation law exists which does {\em not} contain derivatives $D_j P^j,
\;j=m,\ldots,p$ then $0=D_j (CP^j),\;j=1,\ldots,m-1$ is a conservation
law with an arbitrary function $C=C(x^m,\ldots,x^p)$. Reversely,
finding a conservation law involving an arbitrary function
$C(x^m,\ldots,x^p)$ can be exploited to derive a conservation law
involving no derivatives with respect to $x^m,\ldots,x^p$ as it is
described in \cite{Wol99d}. 

\subsection{A faster method to find conservation laws} \label{faster} 
Methods described above decide whether a conservation law can be built
from syzygies, i.e.\ whether there is one in the differential ideal of
the syzygies. Computations to decide this general question are
potentially much more expensive than the other steps of the syzygy
based integration which are all very quick. In the program CRACK
therefore a different, less general but much faster approach is taken.
Instead of determining whether a linear combination of syzygies
exists that makes up a conservation law, the program checks each
individual syzygy whether it can be written as a divergence.

This is done by using conventional integration to integrate the syzygy
with respect to the first variable, say $x$ to obtain $P^x$, then
integrating the remainder with respect to the second variable, say $y$
to obtain $P^y$ and so on. A divergence is obtained when no remainder
remains after the last variable. To find whether the syzygy can be
written as a divergence with only two $P^i$ the above integration is
tried at first with all pairs of two independent variables.
For example, in the case of syzygy  (\ref{e3})
\[ 0 = e_2,_{yzz} - (e_1,_{xx} + e_1,_z)\]
an $x$-integration gives $P^x = - e_1,_x$.
The remainder $e_2,_{yzz} - e_1,_z$ can not be completely
$y$-integrated but $z$-integrated to $P^z = e_2,_{yz} - e_1$.

\section{The redundancy problem} \label{redund} 
Redundant functions are unavoidably generated as soon as 
an equation is conventionally integrated with respect to 
at least two different variables, for example, in the integration of 
$0=A,_{x^1,x^2}$ to $0=A+g(x^1)+h(x^2)$ where $g,h$ depend in addition on all other
independent variables occurring in the expression $A$. 
If $A$ contains $n$ variables $x^1,\ldots,x^n$ then 
the arbitrariness of $g$ and of $h$ overlap to
the extend of one function of $x^3,\ldots,x^n$.
In other words, if $g$ and $h$ are computed from further equations
then there will be one redundant function of $n-2$ variables in the
solution of the original problem.

Let us work out an estimate of how much redundancy is generated when
integrating high order equations. If the conventional method integrates
\[0 = A,_{(x^1)^{m_1},\ldots,(x^n)^{m_n}}\] 
to
\[ A = \sum_{i=1}^n \sum_{j=0}^{m_i-1} g_{ij}\,(x^i)^{\,j} \]
where $g_{ij}$ are free functions of all variables apart from $x^i$
then any two functions $g_{ia}, g_{ib},\; a\neq b$ have no overlap
as their terms $g_{ia}\,(x^i)^{a}, g_{ib}\,(x^i)^{b}$ involve different powers
of $x^i$. Any other pairs of functions $g_{ab}, g_{cd},\; a\neq c$ 
overlap. In total there is an overlap within pairs of functions
$g_{ij}$ equivalent to 
\begin{equation}
\sum_{i=1}^{n-1} \sum_{j=i+1}^n m_i\times m_j          \label{redu}
\end{equation}
functions of $n-2$ variables. In the introductory example
the integration of $0=f,_{yzz}$ gave rise to $1\times 2=2$ redundant
functions of $3-2=1$ variable and in the `real-life' application in
section \ref{reallife} the integration of $0=c_4,_{x_3x_3y_2y_3}$
for $c_4(t,r,x_1,x_2,x_3,y_1,y_2,y_3)$
generates an overlap within pairs of functions equivalent to
$2\times 1 + 2\times 1 + 1\times 1 = 5$ functions of 6 variables and
for $0=c_4,_{x_1x_2x_3x_3x_3y_1y_2y_2}$ even an equivalent of
21 functions of 6 variables. 
The overlap of two functions is partially
also an overlap with other third functions and so on
and should not be counted
twice when trying to account exactly for all the redundancy. But this
correction concerns the arbitrariness content equivalent to 
functions of less than $n-2$
variables so the above formula (\ref{redu}) is a good initial approximation
of redundancy. Keeping in mind that typically a few hundred
such integrations may be necessary, the severity of the problem becomes
obvious. 

{\it Is the redundancy problem an artifact of the chosen examples?} \\
If one determines higher order symmetries of PDEs then the symmetry
conditions may be linear PDEs in, say, 30 independent variables (coordinates
in jet space). Usually the general solution of this overdetermined linear
PDE-system involves constants (corresponding to individual symmetries)
which means that 30 conventional `successive layers' of integrations 
would have to be done, each `layer' containing integrations that
express a function of $n$ variables through functions in $n-1$
variables. In total at least several hundred integrations may become
necessary. From this point of view the above mentioned
application in section \ref{reallife} to compute $c_4$ is typical. 

{\it Could redundancy be prevented otherwise?} \\ Partial
differential equations may contain symmetries involving arbitrary
functions but if not then the general solution of the symmetry
conditions contains only constants. In that case choosing a strictly
lexicographical ordering of derivatives in the elimination process
the differential Gr\"{o}bner basis will involve ordinary differential
equations (ODEs). They may not be in the form of total derivatives but at
least in case they could be integrated, the redundancy problem would
not appear as each ODE is integrated with respect to only one
independent variable. The drawback is that Gr\"{o}bner Basis
computations are well known to be computationally much more expensive
when performed with a lexicographical ordering of variables than when
performed using a total degree ordering of variables. A total degree
ordering will provide shorter equations of lower differential
order but with mixed derivatives, leading to redundancy with
conventional integration. The conclusion is that even in the special
cases where the general solution of the linear PDE system contains
essentially only constants, the syzygy based integration is superior
allowing to use elimination schemes with total degree orderings that
are more efficient than schemes using strictly lexicographical
ordering and still being able to reduce the redundancy problem. 

{\it Does syzygy based integration cure the redundancy problem
completely?} \\ In the course of one syzygy based integration all equations $0=P^i$ are
integrated at once one time. If $0=P^i(e_j)$ is equivalent to the whole system
$0=e_k$, or, like in the introductory example (\ref{e1}),(\ref{e2})
where successive syzygy based integration integrates the system, then
redundancy is avoided. If, on the other hand, only a subsystem of
equations $0=e_k$ is involved in $0=P^i(e_j)$ and the result of a
syzygy based integration has to be substituted in other equations
then redundancy may still appear as recorded in table 1 in section
\ref{threeruns} but to a clearly lesser extend.

{\it Is there another way to determine redundant functions or
constants in order to delete them?} \\
In computations where each free constant in the solution of an
overdetermined PDE-system corresponds to a symmetry or to a
conservation law one is interested to determine and drop redundancy
in order to get an accurate account of their number.
For this purpose a method has been developed
(see \cite{WBM}) but this requires the solution of an overdetermined
PDE-system on its own and may therefore be expensive.

\section{Cases when a syzygy based integration is not useful} \label{useless}
When applying the new integration method to solve a 
PDE-systrem it not only matters whether
all steps are algorithmic but also whether its execution is beneficial. 
Information contained in syzygies is useful if it provides a factorization
of differential operators. If they do not factorize 
(for example, if they are of first order) then a syzygy
based integration can still be useful if more functions are solved for
than new functions are introduced. If the divergence $D_iP^i$
contains more than two derivatives, i.e.\ the conserved 
current $P^i$ has more than 2 components, then the integral equations 
(\ref{over6}) contain
at least one new function $R^{ijk}$ of all variables and we may not gain new
information from the integration if we can not solve for at least 2
functions. This is demonstrated in the following series of 3 examples
with successively more functions to be solve for and an increasing
usefulness of the integration. \vspace{3pt}

{\em Example:}\\
When computing the Gr\"{o}bner basis of the two equations
\begin{eqnarray}
0 & = & f,_x + f,_y \;\;\;\;\;\;  ( =: e_1 )   \label{e17} \\
0 & = & f,_z    \;\;\;\;\;\;\;\;\;\;\;\;\;\;  ( =: e_2 )   \label{e18} 
\end{eqnarray}
for a function $f=f(x,y)$ (and in doing that confirming that they are already a
Gr\"{o}bner basis) one will generate the identity
\begin{equation}
0 = e_2,_x + e_2,_y - e_1,_z.          \label{id10}
\end{equation} 
From identifying $P^x=e_2$ from (\ref{id10}) and the general formula
$P^x=D_yQ^{xy}+D_zQ^{xz}$ together with (\ref{e18}) we identify
$Q^{xy}=0,\, Q^{xz}=f,\, Q^{yz}=f$. With the new function
$R^{xyz}=c(x,y,z)$ substituted into the formula
$Q^{ij}=R^{ijk},_k$ the new equations are
\begin{eqnarray}
0 & = & c,_z         \label{e19} \\
0 & = & f - c,_x     \label{e20} \\ 
0 & = & f + c,_y .    \label{e21}
\end{eqnarray}
After a substitution of $f$ from (\ref{e20}) into (\ref{e21})
they are identical to the original set
(\ref{e17}), (\ref{e18}), only now for a 
function $c$ instead of $f$. No progress was made.
In contrast, for the following two similar examples the integration of
syzygies is advantageous. \vspace{3pt}

{\em Example:}\\
For the equations
\begin{eqnarray}
0 & = & f,_x + g,_y \;\;\;\;\;\;  ( =: e_1 )   \label{e22} \\
0 & = & f,_z    \;\;\;\;\;\;\;\;\;\;\;\;\;\;  ( =: e_2 )   \label{e23} \\
0 & = & g,_z    \;\;\;\;\;\;\;\;\;\;\;\;\;\;  ( =: e_3 )   \label{e24} 
\end{eqnarray}
the identity
\begin{equation}
0 = e_2,_x + e_3,_y - e_1,_z          \label{id11}
\end{equation} 
results. Integrated in the above manner it gives 
\begin{eqnarray}
0 & = &    c,_x  + g      \label{e25} \\
0 & = &  - c,_y  + f      \label{e26} \\
0 & = &    c,_z           \label{e27} 
\end{eqnarray}  
leaving only equation (\ref{e27}) for $c=c(x,y,z)$
to be solved, an improvement compared
to the original system (\ref{e22}) -- (\ref{e24}).
In the next example no equations remain to be solved. \vspace{3pt}

{\em Example:}\\
For the equations
\begin{eqnarray}
0 & = & h,_y - g,_z \;\;\;\;\;\;  ( =: e_1 )   \label{e28} \\
0 & = & f,_z - h,_x \;\;\;\;\;\,  ( =: e_2 )   \label{e29} \\
0 & = & g,_x - f,_y \;\;\;\;\;\;  ( =: e_3 )   \label{e30} 
\end{eqnarray}  
the identity 
\begin{equation}
0 = e_1,_x + e_2,_y + e_3,_z          \label{id12}
\end{equation} 
leads to 
\begin{eqnarray}
0 & = & f + c,_x       \label{e31} \\
0 & = & g + c,_y       \label{e32} \\
0 & = & h + c,_z       \label{e33} 
\end{eqnarray}  
with an arbitrary function $c=c(x,y,z)$ and no remaining equation.

In order to incorporate this method of integration into a general
program for solving over-determined systems the usefulness of integration
has to be judged automatically based on the number of derivatives in
the divergence and the number of functions solved for.
But also other adjustments to the whole program have
to be made. These are discussed in the following short section.

\section{Implementation}  
Apart from the implementation of the algorithm {\sc DivInt} as shown in
the appendix B, also changes to the package {\sc Crack}
were needed in order to automate syzygy
based integrations. When checking integrability conditions
in a Gr\"{o}bner basis computation
the program had to keep track of any resulting identities (syzygies).
This was done in the following way which
is conceptually the same as the extended Buchberger
algorithm (see, for example, the books \cite{BeWe93} and \cite{KrRo00}).

To each equation, for example $e_3$ in (\ref{e3a}),
we will assign not only a value, like $f,_{yzzz}$,
but also, what we will call a `history-value' or short
`history', i.e.\ $e_2,_{yzz} - e_1,_{xx}$. 
This history of an equation expresses one equation in terms of
other equations, i.e.\ how it was historically computed doing the 
algebraic or differential Gr\"{o}bner basis computation.
At the beginning the history of each equation $e_a$ is $e_a$ itself. 
Whenever a new equation is computed then not only its value but also
its history is calculated. For example, when in this
example $f,_{yzzz}$ is eliminated from equation (\ref{e3a}) using
equation (\ref{e1}) 
then a new equation $0=e_4$ is generated where $e_4$ has the value 0
(as all terms cancel) and has the history value $e_3-e_1,_z$ where $e_3$
and $e_1$ are replaced by their history values. 
The history of $e_1$ is $e_1$ whereas the history of $e_3$
is $e_2,_{yzz} - e_1,_{xx}$ giving for $e_4$ the history
$e_2,_{yzz} - e_1,_{xx} - e_1,_z$ as is shown in (\ref{e3}).

In the next section a substantial application is described which is suitable
to demonstrate the advantages of the new integration method.

\section{The application that led to the development of 
         the syzygy based integration} \label{reallife} 
\subsection{The problem}
A problem introduced to the author by Stephen Anco 
concerns the computation of all
conservation laws of the radial SU(2) chiral equation 
in 2 spatial dimensions where the integrating
factors are of at most $2^{\mbox{\scriptsize nd}}$ order.
The equation can be 
written as a first order system for two 3-component vectors 
\mbox{{\bf j}}(r,t), \mbox{{\bf k}}(r,t):
\begin{eqnarray}
\mbox{{\bf k}},_t & = & \mbox{{\bf j}},_r + \;
                        \mbox{{\bf j}} \times \mbox{{\bf k}} \label{SU2-1}\\
\mbox{{\bf j}},_t & = & (r\mbox{{\bf k}}),_r/r .             \label{SU2-2}
\end{eqnarray}
Equation (\ref{SU2-2}) is already in conservation law form:
\[(r\mbox{{\bf j}}),_t + (-r\mbox{{\bf k}}),_r = 0            \]
and the only other known conservation law (of energy) has zeroth order
integrating factors: 
\begin{equation}
r\mbox{{\bf k}}\cdot \left[\mbox{{\bf k}},_t - \mbox{{\bf j}},_r
-\mbox{{\bf j}} \times \mbox{{\bf k}}\right]
+ \mbox{{\bf j}}\cdot \left[\mbox{{\bf j}},_t - (r\mbox{{\bf k}}),_r/r\right]\\
= \left(\frac{r}{2}\left(\mbox{{\bf j}}\cdot \mbox{{\bf j}} + \mbox{{\bf k}}\cdot
\mbox{{\bf k}}\right)\right),_t + 
\left(-r \mbox{{\bf j}}\cdot \mbox{{\bf k}}\right),_r=0  \label{SU2-3}
\end{equation}
The existence conditions for conservation laws below were
generated with the program {\sc ConLaw2} described in \cite{Wol99a}.
It generates conditions for 6 integrating factors $Q_1, \ldots, Q_6$
(like the multipliers $rk_1, rk_2, rk_3, j_1, j_2, j_3$ 
on the left hand side of (\ref{SU2-3})).
Each of the $Q_i$ is an unknown function of 20 independent variables
$t,r,
 \mbox{\bf j}, 
 \mbox{\bf k}, 
 \mbox{\bf l}\; (=\mbox{\bf j},_r), 
 \mbox{\bf m}\; (=\mbox{\bf k},_r),
 \mbox{\bf u}\; (=\mbox{\bf j},_{rr}), 
 \mbox{\bf w}\; (=\mbox{\bf k},_{rr})$.
The system consists of 18 conditions of the form
\[0 = Q_1,_{u_1} - Q_4,_{w_1}r \]
and 6 conditions of the form
\footnotesize
\begin{eqnarray*}
0\!\!\!&=\!\!\!&Q_3,_{j_1}l_1r^2 + Q_3,_{l_1}u_1r^2 + Q_3,_{j_2}l_2r^2 + 
Q_3,_{l_2}u_2r^2 + Q_3,_{j_3}l_3r^2 + Q_3,_{l_3}u_3r^2 
+ Q_3,_{k_1}m_1r^2 + Q_3,_{m_1}w_1r^2 
\\ & &
+ Q_3,_{k_2}m_2r^2 + Q_3,_{m_2}w_2r^2 
+ Q_3,_{k_3}m_3r^2 + Q_3,_{m_3}w_3r^2 
+ Q_3,_{r}r^2 - Q_6,_{j_1}k_1r^2 
- Q_6,_{j_1}m_1r^3 + Q_6,_{l_1}k_1r 
\\ & &
- Q_6,_{l_1}m_1r^2 - Q_6,_{l_1}w_1r^3 
- 2Q_6,_{u_1}k_1 + 2Q_6,_{u_1}m_1r 
- Q_6,_{u_1}w_1r^2 - Q_6,_{j_2}k_2r^2 
- Q_6,_{j_2}m_2r^3 + Q_6,_{l_2}k_2r 
\\ & &
- Q_6,_{l_2}m_2r^2 - Q_6,_{l_2}w_2r^3 
- 2Q_6,_{u_2}k_2 + 2Q_6,_{u_2}m_2r 
- Q_6,_{u_2}w_2r^2 - Q_6,_{j_3}k_3r^2 
- Q_6,_{j_3}m_3r^3 + Q_6,_{l_3}k_3r 
\\ & &
- Q_6,_{l_3}m_3r^2 - Q_6,_{l_3}w_3r^3 
- 2Q_6,_{u_3}k_3 + 2Q_6,_{u_3}m_3r 
- Q_6,_{u_3}w_3r^2 - Q_6,_{k_1}l_1r^3 
- Q_6,_{k_1}j_2k_3r^3 + Q_6,_{k_1}j_3k_2r^3 
\\ & &
- Q_6,_{m_1}u_1r^3 - Q_6,_{m_1}j_2m_3r^3 
- Q_6,_{m_1}l_2k_3r^3 + Q_6,_{m_1}j_3m_2r^3 
+ Q_6,_{m_1}l_3k_2r^3 - Q_6,_{w_1}j_2w_3r^3 
- 2Q_6,_{w_1}l_2m_3r^3 
\\ & &
- Q_6,_{w_1}u_2k_3r^3 
+ Q_6,_{w_1}j_3w_2r^3 + 2Q_6,_{w_1}l_3m_2r^3 
+ Q_6,_{w_1}u_3k_2r^3 + Q_6,_{k_2}j_1k_3r^3 
- Q_6,_{k_2}l_2r^3 - Q_6,_{k_2}j_3k_1r^3 
\\ & &
+ Q_6,_{m_2}j_1m_3r^3 + Q_6,_{m_2}l_1k_3r^3 
- Q_6,_{m_2}u_2r^3 - Q_6,_{m_2}j_3m_1r^3 
- Q_6,_{m_2}l_3k_1r^3 + Q_6,_{w_2}j_1w_3r^3 
+ 2Q_6,_{w_2}l_1m_3r^3 
\\ & &
+ Q_6,_{w_2}u_1k_3r^3 
- Q_6,_{w_2}j_3w_1r^3 - 2Q_6,_{w_2}l_3m_1r^3 
- Q_6,_{w_2}u_3k_1r^3 - Q_6,_{k_3}j_1k_2r^3 
+ Q_6,_{k_3}j_2k_1r^3 - Q_6,_{k_3}l_3r^3 
\\ & &
- Q_6,_{m_3}j_1m_2r^3 - Q_6,_{m_3}l_1k_2r^3 
+ Q_6,_{m_3}j_2m_1r^3 + Q_6,_{m_3}l_2k_1r^3 
- Q_6,_{m_3}u_3r^3 - Q_6,_{w_3}j_1w_2 r^3 
- 2Q_6,_{w_3}l_1m_2r^3 
\\ & &
- Q_6,_{w_3}u_1k_2r^3 
+ Q_6,_{w_3}j_2w_1 r^3 + 2Q_6,_{w_3}l_2m_1r^3 
+ Q_6,_{w_3}u_2k_1r^3 - Q_6,_{t}r^3 - k_1Q_2r^2 
+ k_2Q_1r^2 
\end{eqnarray*}
\normalsize

After introducing new unknown functions $x_i,y_i$ through
$u_i=x_i+y_i, w_i=x_i-y_i$ the 18 short equations took the
form of a total derivative and each one could be integrated on its
own but when the computed functions were substituted only indirectly 
separable equations (ISEs) like (\ref{e9}) were obtained.\footnote{
Although each of the ISEs is over-determined on its own,
this over-determination can not be utilized easily because there is no
independent variable which occurs only explicitly that would lead to direct
separations.}

Despite of the initial success in performing these 
integrations all attempts to complete the solution of the
over-determined system failed with the 1999 version of {\sc Crack}.
That this was not simply a matter of lacking computing power became
obvious after extracting a small sub-system of equations for only one of
the unknown functions\footnote{New constants and functions of 
integration are all called $c_i$ 
in {\sc Crack} with successively increasing subscript.}
$\;c_4(t,r,x_1,x_2,x_3,y_1,y_2,y_3)$
where some of the equations are easy to integrate:
\begin{eqnarray}
0&=&c_4,_{x_3x_3y_2y_3} = c_4,_{x_1x_2y_1y_3y_3} =
c_4,_{x_1x_2y_1y_1y_3} = c_4,_{x_1x_2x_3y_1y_1} = 
c_4,_{x_2x_3x_3x_3y_1y_1y_3} \nonumber \\
 &=&c_4,_{x_1x_2x_3x_3x_3y_1y_2y_2} 
= c_4,_{x_1x_2x_2x_3y_1y_1y_2y_2} 
= c_4,_{x_1x_2x_3x_3y_3y_3y_3} -  
c_4,_{x_2x_3x_3x_3y_1y_3y_3}  \nonumber \\
 &=&c_4,_{x_1x_2x_3x_3y_1y_2y_2} - 
   2c_4,_{x_1x_2x_2x_3y_1y_2y_3} = c_4,_{x_1x_2x_3x_3y_1y_2} - 
  2c_4,_{x_1x_2x_2x_3y_1y_3} - 
    c_4,_{x_1x_2x_2x_3x_3y_1y_3}x_3  \nonumber \\
 &=&c_4,_{x_1x_2x_3x_3x_3y_1y_3}x_1 - 
    c_4,_{x_1x_2x_3x_3y_3y_3} + 
    c_4,_{x_2x_3x_3x_3y_1y_3}  \nonumber \\
 &=&c_4,_{x_3x_3x_3y_1y_3}x_1 + 
    c_4,_{x_3x_3y_1y_3y_3}y_1 - 
    c_4,_{x_3x_3y_3y_3}       \label{c4sys} \\
 &=&c_4,_{x_1x_2x_3x_3y_1y_2y_2y_2}y_3 + 
   2c_4,_{x_1x_2x_2x_2y_1y_2y_3} - 
   2c_4,_{x_1x_2x_2x_3y_1y_2y_2} +
    c_4,_{x_1x_2x_2x_3x_3y_1y_2y_2}x_3  \nonumber \\
 &=&c_4,_{x_1x_2x_3x_3y_1y_2y_2}y_3 + 
  2c_4,_{x_1x_2x_2x_2x_3y_1y_3}x_3 + 
  2c_4,_{x_1x_2x_2x_2y_1y_3} - 
  2c_4,_{x_1x_2x_2x_3y_1y_2}  \nonumber \\
 &=&c_4,_{x_1x_2x_3x_3x_3y_1y_2}x_1x_3 - 
  3c_4,_{x_1x_2x_3x_3y_1y_2}x_1 + 
  6c_4,_{x_1x_2x_2x_3y_1y_3}x_1  \nonumber \\
& & - c_4,_{x_1x_2x_2x_3x_3y_3y_3}x_3^2 +
    c_4,_{x_2x_2x_3x_3x_3y_1y_3}x_3^2   \nonumber 
\end{eqnarray}
Even the solution or at least simplification of this sub-system was
not possible. The problem was not to find equations with the form of a
total derivative and to integrate them. The problem was the growing
number of new functions of integration (which did still depend on 
7 variables) and the appearance of too many only indirectly
separable equations (ISEs). 

Since 1999 the module for handling 
ISEs has been improved considerably. 
The current version of {\sc Crack} (Dec.\ 2001)
can simplify the above system quickly using the conventional
integration of total derivatives.
Nevertheless, by adding the ability of performing syzygy based
integrations the computation speeds up further and the solution 
involves fewer redundant arbitrary functions. Tests described below
show that syzygy based integrations are well suited to be performed
along the computation of a differential Gr\"{o}bner basis
without the negative side effect of introducing too many redundant
functions. By that Gr\"{o}bner basis computations 
can be cut short and the risk of a memory explosion be lowered.

\subsection{A comparison of three computer runs} \label{threeruns}
Before describing the details of 3 different computer runs, 
a few comments about the setup have to be made. The package 
{\sc Crack} for solving and simplifying over-determined
PDE-systems contains about 30 modules for different actions 
to be taken either with individual equations or with
groups of equations of the system. Modules used to solve systems like 
(\ref{c4sys}) are
\begin{enumerate}
\item 
     Direct separation of an equation with respect to some variable that occurs
     only explicitly in the equation.
\item 
     Substitution of a function $f$ either by zero or by at most 2
     terms and only if other functions occurring in these 2 terms
     depend on fewer variables than $f$.
\item 
     Integration of an equation if it consists of a single derivative
     with respect to only one variable.
\item 
     Elimination of a function $f$ from any equation if $f$ occurs only
     algebraically and linearly and if $f$ depends on
     all variables occurring in this equation. Substitution of $f$
     in all other equations.
\item 
     Deleting of any redundant equations as described on the bottom of
     the overview in section \ref{general}.
\item 
     Integration based on a syzygy in conservation law form.
\item 
     Conventional integration of a PDE but only if sufficiently many
     integrations are possible such that the integrated equation can be
     used for a substitution.
\item 
     Indirect separation of an equation (ISE). (This is a complex step
     which can invoke other direct separations and indirect separations of
     resulting equations.)
\item 
     Reduction of the leading derivative of one equation with the help
     of another equation or formulation of an integrability condition
     between two equations. (This is a typical step in a Gr\"{o}bner
     basis computation.)
\item 
     Any integration of any equation even if not complete.
\end{enumerate}
These modules are called in a specific sequence which can be chosen
by specifying a list of numbers, each number representing one module.
For example, if in table 1, column 2 the priority list of run 1 is chosen to
be 1 2 3 4 8 9 7 10 then the modules as numbered above
are tried in this order until one
module is successful and then they are again tried beginning with 1
and so on. This is only a simplified description of the operation 
of {\sc Crack} but it is sufficient for the purpose of this section.
\vspace*{6pt}

\small
\hspace*{-3mm}\begin{tabular}{|c|l|c|c|c|c|c|c|c|}\hline
run&priority list&\# of&time&\# of terms &\multicolumn{4}{c|}{\# of redundant functions} \\ \cline{6-9}
 &of actions&steps  &in sec& in equ. (\ref{remainDE})&of 6 var.&of 5 var.&of 4 var.&of 3 var. \\ \hline
1&1 2 3 4 8 9 7 10    & 1077&   124&     6       &    7      &    16     &    2    &    0     \\ \hline
2&1 2 3 4 7 8 9 10    & 1175&   122&    12       &    4      &    45     &   23    &    5     \\ \hline
3&1 2 3 4 5 8 6 9 7 10&  362&    23&     8       &    2      &    19     &    2    &    0     \\ \hline
\end{tabular}
\begin{center}
Table 1. A comparison of three different runs on the system
(\ref{c4sys}). 
\end{center}
\normalsize

In table 1 three computer runs are compared.
Column 3 gives the number of successful calls of the
modules in the priority lists. Times shown in column 4 have been measured in a session
of the computer algebra system REDUCE version 3.7 with 120 MByte memory
(although only a few MByte are needed for this computation) on a 1.7
GHz PC Pentium 4 under Linux. 
Column 5 gives the number of terms in the single
unsolved equation which in the solution (\ref{minSOL}) below is the
equation (\ref{remainDE}). In the remaining 4 columns the number 
of redundant functions of 6, 5, 4, or 3 variables is shown. For example,
if two functions $f(x,y,z)$ and $zg(x)$ occur always together such that a
substitution $f+zg \rightarrow f$ has the same effect as 
$g \rightarrow 0$ then $g$ can be set to zero without loss of
generality. 

\subsection{Conclusions from the test} \label{conclu}
The central issue in these runs is, whether integrations
(modules 6 and 7) are given a higher priority than the formulation of
integrability conditions (module 9) or a lower priority.
If integrability conditions have a higher priority than
integrations, as in run 1, then 
at first a complete differential Gr\"{o}bner basis is computed before
integrations start. The benefit is that the 
differential order of equations is as
low as possible when integrations start (assuming a total degree
ordering is used in the differential Gr\"{o}bner basis computation). 
Consequently fewer integrations are necessary and
fewer functions will be generated which turn out later to
be redundant. The
disadvantage is that the computation of integrability conditions may
take very long and blow up the systems size, or may even be
practically impossible. 

One can attempt to give integrations a higher priority at
the price of more redundant functions in the solution. 
This was done in run 2. The benefit may
be considerable, only in our small system (\ref{c4sys})
the Gr\"{o}bner basis computation
is not expensive at all, so the advantage of early integrations does
not become obvious here. But the disadvantage becomes obvious.
Integrating higher order equations generates more new functions with
many of them turning out to be redundant at the end.

Finally, in the third run we get the best of
both previous runs. Here, early integrations use syzygies in
conservation law form as soon as they become available. The lowered
differential order of equations reduces the complexity of the
remaining Gr\"{o}bner basis computation.
Also, because with each integration at least 2 equations $0=P^i$ are 
satisfied, the number of new functions of integration is low
and the number of variables these functions depend on is reduced.
Consequently, only few functions turn out to be redundant in the
computed solution as seen in columns 6-9 of table 1.

The following solution is obtained in run 3 after redundant functions
have been deleted (by hand) leaving 11 functions of 6
variables, 8 functions of 5 variables and 2 functions of 4 variables.
It is equivalent to the solutions returned in runs 1 and 2.
\begin{eqnarray}
c_4&=&c_{100},_{x_2}x_3y_1y_2 
+ \mbox{\small $\frac{1}{2}$}c_{100},_{x_3}x_1y_3^2 
+ c_{125},{x_2}x_3^2y_1 + c_{125},_{y_2}x_3y_1y_3 
+ c_{133},_{x_2}x_3y_1 + c_{133},_{y_2}y_1y_3  \nonumber \\
& & + c_{213},_{x_3}x_1 
+ c_{213},_{y_3}y_1 + c_{100}y_1y_3 - c_{109}x_2x_3y_1 
+ c_{170} + c_{172} + c_{173}x_3 + c_{181}y_3 + c_{191} \label{minSOL} \\
&&- c_{192} - c_{193}x_3 - c_{194} + c_{200} + c_{205} 
- \mbox{\small $\frac{1}{2}$}c_{65}x_3^2y_3 - c_{81}x_3 - c_{83}  \nonumber
\end{eqnarray}
All functions depend on $t,r$ and in addition on further variables in
the following way:
\[\begin{array}{l}
c_{83}(x_2,x_3,y_1,y_2),\;\; c_{81}(x_2,y_1,y_2,y_3),\;\; c_{173}(x_1,x_2,y_2,y_3),\;\;
c_{172}(x_1,x_2,y_2,y_3),\;\; c_{170}(x_1,x_2,x_3,y_2),\;\; \\
c_{194}(x_1,x_3,y_1,y_3),\;\; c_{193}(x_1,y_1,y_2,y_3),\;\;
c_{192}(x_1,y_1,y_2,y_3),\;\; c_{191}(x_1,x_3,y_1,y_2),\;\; 
c_{205}(x_2,y_1,y_2,y_3),\;\; \\
c_{200}(x_1,x_2,y_1,y_2),\;\; c_{100}(x_1,x_2,x_3),\;\; 
c_{125}(x_1,x_2,y_2),\;\; c_{133}(x_1,x_2,y_2),\;\; c_{181}(x_1,x_2,x_3),\;\; \\
 c_{213}(x_2,x_3,y_3),\;\; c_{230}(x_1,y_1,y_3),\;\; c_{229}(x_1,y_1,y_3),\;\;
c_{228}(x_1,x_3,y_1),\;\; c_{65}(x_2,y_1),\;\; c_{109}(x_1,y_2).
\end{array}\]
The function $c_{194}$ has to satisfy the condition
\begin{equation}
0=c_{194},_{x_3y_1}x_1 + c_{194},_{y_1y_3}y_1 - c_{194},_{y_3} 
 - c_{228} - c_{229} - c_{230}x_3 ,     \label{remainDE}
\end{equation}
all other functions are free.
The result of the conservation law investigation for the SU(2) chiral
equation in the form (\ref{SU2-1}), (\ref{SU2-2}) is that no other
conservation laws with integrating factors of at most 
$2^{\mbox{\scriptsize nd}}$ order exist.

More remarks concerning the collaboration of modules:
\begin{itemize}
\item Syzygy based integration can not replace conventional
integration. If equations become decoupled then no integrability
conditions apply and the equations have to be integrated conventionally if
possible. 
\item The usefulness of
conventional integration relies very much on the efficiency
of a module for the indirect separation (module 8 in the above list).
The corresponding implementation in {\sc Crack} will be described elsewhere.
\item The issue of avoiding redundant functions is serious when a system
like (\ref{c4sys}) is only a subsystem of a larger system and the
solution of the smaller system is to be substituted in the larger one. 
Redundant functions would complicate the solution of the larger system
unnecessarily. On the other hand, the identification and deletion 
of redundant functions using a method described in \cite{WBM}, is
difficult and may be more expensive than the solution/simplification of
the system itself. This method does not prevent redundancy, it only can
identify it in the solution.
\end{itemize}

The package {\sc Crack} is distributed together with the computer
algebra system REDUCE. A newer version can be down-loaded from
{\tt http://lie.math.brocku.ca/twolf/crack}.

\section{Summary}
An integration method has been proposed that is applicable for linear
PDE-systems that admit syzygies, i.e.\ systems which are
overdetermined as a whole or contain an overdetermined subsystem.
It therefore can not replace the straight forward integration of exact
PDEs but when applicable it has a number of advantages:
\begin{itemize}
\item The information on which the integration is based is taken from
  syzygies in conservation law form. Syzygies are a by-product of the
  computation of differential Gr\"{o}bner Basis.
\item Because not a single equation is integrated but a number of
  equations $(0=P^i)$ at once, fewer functions of integration, depending on
  fewer variables are introduced in the process. 
\item The problem of conventional integration to introduce redundant
  functions when integrating with respect to different variables is
  either prevented or significantly reduced.
\item The new integration produces apart from integrated equations
  also new syzygies which are often the basis for continuing the
  integration further without having to compute new
  syzygies through a new Gr\"{o}bner basis computation.
\item Syzygy based integration, conventional integration and
  elimination complement one another well in solving overdetermined
  linear PDE-systems if given the right priorities.
\end{itemize}

\newpage
\section*{Appendix A: Continuation of the introductory example} 
In this appendix we continue the introductory example by performing
three more syzygy based integration steps. The computation is broken
up into items.
The number(s) at the start of each item refer to the line number
of the corresponding step in the overview at the end of section
\ref{general}. 
\begin{description}
\item[(\ref{over1}):] 
The remaining system to solve consists of 
\begin{eqnarray*}
0 = f,_{xx} + f,_z     \;\;\;\;\;\;     ( = e_2 ) \\
0 = f,_{xyz} - c_1 .   \;\;\;\;\;\;    ( = e_4 )  
\end{eqnarray*}
\item[(\ref{over2}),(\ref{over3}):] 
satisfying the identity in conservation law form 
\[ 0 = (- e_4,_x),_x \, + \, (e_2,_{xy} - e_4),_z \]
\item[(\ref{over3a}):] 
with only 2 derivatives.
\item[(\ref{over4}):] 
Proceeding as in the first integration step we now identify as the conserved current
\begin{eqnarray}
\hat{P}^x & = & - e_4,_x = - f,_{xxyz} = - \hat{Q},_z      \label{id7a} \\
\hat{P}^z & = &   e_2,_{xy} - e_4 = f,_{xxxy} + c_1 = \hat{Q},_x  \label{id8a}
\end{eqnarray}
\item[(\ref{over5}):] 
and as the new potential $\hat{Q}$ we either identify or compute using
algorithm {\sc DivInt} in appendix B
\[\hat{Q} = f,_{xxy} + x c_1  \]
\item[(\ref{over6}),(\ref{over7}):] 
giving the new equation 
\begin{equation}
0 = \hat{Q} -c_2 = f,_{xxy} + x c_1 - c_2 \;\;\;\;\;\; ( =: e_5 ) \label{e5}
\end{equation} 
with the new function of integration $c_2=c_2(y)$. 
\item[(\ref{over8}),(\ref{over9}):] 
Equation $e_4$ is redundant as it turns up purely algebraically in 
\[ 0=\hat{P}^z- \hat{Q},_x =  e_2,_{xy} - e_4 -  e_5,_x. \]
\item[(\ref{over9a}):] 
Substitution of $e_4$ in (\ref{id7a}) gives the new identity 
\begin{equation}
   0 = - e_2,_{xxy} + e_5,_{xx} + e_5,_z.               \label{id9a}
\end{equation} 
\item[(\ref{over4}):] 
This is as well a divergence with only two terms
\begin{eqnarray}
\bar{P}^x & = & - e_2,_{xy} + e_5,_x = - f,_{xyz} + c_1 = - \bar{Q},_z  \label{id7b} \\
\bar{P}^z & = &   e_5 = f,_{xxy} + xc_1 - c_2 = \bar{Q},_x  \label{id8b}
\end{eqnarray}
\item[(\ref{over5}):] 
and the new potential $\bar{Q}$
\[\bar{Q} = f,_{xy} + \frac{x^2}{2} c_1 - xc_2 - zc_1 \]
\item[(\ref{over6}),(\ref{over7}):] 
giving the new equation 
\begin{equation}
0 = \bar{Q} -c_3 = f,_{xy} + \frac{x^2}{2} c_1 - xc_2 - zc_1 - c_3
\;\;\;\;\;\; ( =: e_6 ) \label{e6}
\end{equation} 
with the new function of integration $c_3=c_3(y)$. 
\item[(\ref{over8}),(\ref{over9}):] 
Now, equation $e_5$
is redundant as it turns up purely algebraically in 
\[ 0=\bar{P}^z- \bar{Q},_x =  e_5 - e_6,_x. \]
\item[(\ref{over9a}):] 
Substitution of $e_5$ in (\ref{id7b}) gives the new identity 
\begin{equation}
   0 = - e_2,_{xy} + e_6,_{xx} + e_6,_z.               \label{id9b}
\end{equation} 
\item[(\ref{over4}):] 
This is a divergence as well and we will perform the integration
cycle one more time with
\begin{eqnarray}
\check{P}^x & = & - e_2,_y + e_6,_x = - f,_{yz} + xc_1 - c_2 = - \check{Q},_z  \label{id7c} \\
\check{P}^z & = &   e_6 = f,_{xy} + \frac{x^2}{2} c_1 - xc_2 - zc_1 - c_3 = \check{Q},_x  \label{id8c}
\end{eqnarray}
\item[(\ref{over5}):] 
and the new potential $\check{Q}$
\[\check{Q} = f,_y + \frac{x^3}{6} c_1 - \frac{x^2}{2}c_2 - xzc_1 + zc_2 - xc_3\]
\item[(\ref{over6}),(\ref{over7}):] 
giving the new equation 
\begin{equation}
0 = \check{Q} - c_4 = f,_y + \frac{x^3}{6} c_1 - \frac{x^2}{2}c_2 - xzc_1 + zc_2 - xc_3 - c_4
\;\;\;\;\;\; ( =: e_7 ) \label{e7}
\end{equation} 
with the new function of integration $c_4=c_4(y)$. 
\item[(\ref{over8}),(\ref{over9}):] 
Now, equation $e_6$
is redundant as it turns up purely algebraically in $\check{P}^z$ in  (\ref{id8c})
\[ 0=\check{P}^z- \check{Q},_x =  e_6 - e_7,_x. \]
\item[(\ref{over9a}):] 
Substitution of $e_6$ in (\ref{id7c}) gives the new identity 
\begin{equation}
   0 = - e_2,_y + e_7,_{xx} + e_7,_z.               \label{id9c}
\end{equation} 
\end{description}
The conclusion of this example is shown in section \ref{intro1} below
equation (\ref{id9c2}). As argued there the syzygy based integration
of equation (\ref{id9c}) is not advantageous as (\ref{id9c}) has a
conservation law form with 3 derivatives instead of two. Instead one
rather integrates (\ref{e7}) with respect to $y$ 
and substitutes $f$ in the remaining
equation (\ref{e2}).

\newpage
\section*{Appendix B: The algorithm {\sc DivInt}}  
The following algorithm computes expressions 
$Q^{ij}(x^n,f^\alpha_J)=Q^{[ij]}$
that satisfy $ D_j Q^{ij}=P^i$. The given
$P^i=P^i(x^n,f^\alpha_J)$ are assumed to satisfy $D_i P^i = 0$ identically
in all $f^\alpha_J$.

\begin{tabbing}
\noindent  
\ \ 1\ \ \ \= {\bf Algorithm} {\sc DivInt}  \\
\ \ 2      \> {\bf Input} \hspace{8pt} variables: $x^n$,
              functions: $f^\alpha$ and conserved current:
              $P^i=P^i(x^n,f^\alpha_J)$ \\
\ \ 3      \> {\bf Output} \= $Q^{ij}(x^n,f^\alpha_J),\;j>i$  
              \hspace{8 pt} \% satisfying
              $ D_j Q^{ij}=P^i$,  \\
\ \ 4      \>           \> $E, F\;\;$ \hspace{68pt} \% 
                           $E:$ list of new additional equations \\
\ \ 5      \>           \> \hspace{103pt} \% $F:$ list of new
           additional functions \\
\ \ 6      \> {\bf Body}  \hspace{115pt} \% no summation over double
                                            indices below\\
\ \ 7      \> \ \ \ \= $E:=\{\},\; F:=\{\},\;
                        Q^{ij}:=0,\;\, \;\;
\mbox{with}\;\;\;i,j \in 1,\ldots,p,\; j>i$\\
\ \ 8      \>       \>  \\
\ \ 9      \>       \> \% Integrate all terms with functions $f^\alpha$
                          depending on all variables \\
 10        \>       \> {\bf for} $i:=1$ to $(p-1)$ {\bf do} \\
 11        \>       \> \ \ \ \= {\bf for} $j:=i+1$ to $p$ {\bf do} \\
           \>       \>       \> \ \ \ \=   \kill
 12        \>       \>       \>       \> {\bf while} $P^i$ contains a term
                                          $a^{iJ}\partial_jf^\alpha_J$
       {\bf do} \hspace{20pt} \% i.e.\ while any derivative of any $f^\alpha$ \\
 13        \>       \>       \>       \> \hspace{217pt} \% occurs that involves $\partial_{j}$ \\
           \>       \>       \>       \> \ \ \ \=   \kill
 14        \>       \>       \>       \>       \>
       $P^i \rightarrow P^i - D_j(a^{iJ}f^\alpha_J)$  \\
 15        \>       \>       \>       \>       \>
       $P^j \rightarrow P^j + D_i(a^{iJ}f^\alpha_J)$  \\
 16        \>       \>       \>       \>       \>
       $Q^{ij} \rightarrow Q^{ij} + a^{iJ}f^\alpha_J$ \\
 17        \>       \>  \\
 18        \>       \> \% Integrate all derivatives involving
                        functions $f^\alpha$ not depending on all variables \\
 19        \>       \> {\bf for} $i:=2$ to $p$ {\bf do} \\
 20        \>       \> \ \ \ \= {\bf for} $j:=1$ to $i-1$ {\bf do} \\
           \>       \>       \> \ \ \ \=   \kill
 21        \>       \>       \>       \> {\bf while} $P^i$ contains a term
                                          $a^{iJ}\partial_jf^\alpha_J$
       {\bf do} \hspace{20pt} \% i.e.\ while any derivative of any $f^\alpha$ \\
 22        \>       \>       \>       \> \hspace{217pt} \% occurs that involves $\partial_{j}$ \\
           \>       \>       \>       \> \ \ \ \=   \kill
 23        \>       \>       \>       \>       \>
       $P^i \rightarrow P^i - D_j(a^{iJ}f^\alpha_J)$  \\
 24        \>       \>       \>       \>       \>
       $Q^{ji} \rightarrow Q^{ji} - a^{iJ}f^\alpha_J$ \\
 25        \>       \>  \\
 26        \>       \> \% Integrate remaining terms \\
 27      \>       \> {\bf for} $i:=1$ to $p$ {\bf do} \\
 28      \>       \> \ \ \ \= {\bf if} $P^i\neq 0$ {\bf then} \\
 29      \>       \>       \> \ \ \ \= \% integrate each term 
       $a^{iJ}f^\alpha_J$ of $P^i$ with respect to any one $x^j\neq x^i$ \\
 30      \>       \>       \>       \> 
       \% preferably one $x^j$ with $\partial_j f^\alpha=0$
       in the following way: \\
 31      \>       \>       \>       \> {\bf if} $\partial_j f^\alpha=0$ {\bf then} \\
 32      \>       \>       \>       \> \ \ \ \=
                                         $q:=f^\alpha_J\int a^{iJ}\,dx^j$ \\
 33      \>       \>       \>       \>       \>
       $P^i \rightarrow P^i - D_j q$  \\
 34      \>       \>       \>       \>       \>
         {\bf if} $j>i$ \= {\bf then}  \= $Q^{ij} \rightarrow Q^{ij} + q$ \\
 35      \>       \>       \>       \>       \>
                  \> {\bf else}  \> $Q^{ji} \rightarrow Q^{ji} - q$ \\
 36      \>       \>       \>       \> {\bf else} \\
 37      \>       \>       \>       \> \ \ \ \= Introduce a new
function $f^\beta(x^1,\ldots,x^{i-1},x^{i+1},\ldots,x^{p})$ \\
 38      \>       \>       \>       \>       \> $F \rightarrow F \cup \{f^\beta\}$ \\
 39      \>       \>       \>       \>       \> $E \rightarrow E \cup
\{0=\partial_j f^\beta - a^{iJ}f^\alpha_J\}$ \\
 40      \>       \>       \>       \>       \>
       $P^i \rightarrow P^i - a^{iJ}f^\alpha_J$  \\
 41      \>       \>       \>       \>       \>
       {\bf if} $j>i$ \= {\bf then} \= $Q^{ij} \rightarrow Q^{ij} + f^\beta$ \\
 42      \>       \>       \>       \>       \>
                \> {\bf else} \> $Q^{ji} \rightarrow Q^{ji} - f^\beta$ \\
 43      \>       \>   {\bf return} $Q^{ij}(x^n,f^\alpha_J)$,
                                      $E$ (list of new equations),
                                      $F$ (list of new functions) 
\end{tabbing}

\newpage
{\bf Explanation of the algorithm} \vspace{3pt}

{\em Lines 9 - 16} \\
This part of the procedure is sufficient if the
input expressions $P^i(x^n,f^\alpha_J)$ do only contain functions
$f^\alpha$ depending on all $p$ independent variables $x^1,\ldots,x^p$.

A typical example:
If an expression $P^y$ contains a term $f,_z$
then $D_y P^y$ (no summation) contains $\partial_y f,_z$
which has to be cancelled by $ - \partial_z f,_y$ from 
$D_z P^z$ (no summation) to give $0= D_k P^k$ (summation)
identically in all $f_J$. This means $P^z$ 
contains $ - f,_y$. In this short example
the lines 14 - 16 would subtract $f,_z$ from $P^y$, 
subtract $-f,_y$ from $P^z$ and add $f$ to $Q^{yz}$.
There is no principal difference between $P^y$ containing a term $f,_z$
or $P^y$ containing $a^{iJ}\partial_zf^\alpha_J$.

As both, $P^i$ and $P^j$ are updated in lines 14 and 15, $j$ does not
run over indices $1 \ldots i-1$.
Because $Q^{ii}=0$ ($Q^{ij}$ is antisymmetric) there is no need to
integrate an $i$-derivative in $P^i$ and therefore $j$ starts from
$i+1$ in line 11.

If all terms in all $P^i$ contain a function $f^\alpha$ of all
variables then any term in any $Q^{ij}$ occurs twice, once with
an $x^j$-derivative in $P^i$ and once as negative $x^i$-derivative
in $P^j$. When the program completed lines 10 - 16, all $P^i$
have the value zero and the solution $Q^{ij}$ is found (for $i < j$,
values for $Q^{ji}$ follow from the antisymmetry). \vspace{3pt}

{\em Lines 18 - 42} \\
The only possibility that after completing lines 10 - 16 not all $P^i$
are already zero occurs if some $f^\alpha$ do not depend on all
variables. That is, for example, the case if functions entered the problem due to
running {\sc DivInt} previously in earlier integrations.
In general, if terms remain in some $P^i$ which necessarily depend on
less than all variables then one can always complete
the integrations by introducing new functions (collected in a list $F$ in
line 38) that have to satisfy additional equations (collected in a list $E$
in line 39). In order to minimize the number of additional functions
and additional equations the lines 19 - 24 integrate terms that are
$x^j$-derivatives in $P^i$ ($j \neq i$)
and lines 31 - 35 integrate terms by changing the explicit appearance
of $x^j$. This is shown in the following examples. \vspace{3pt}

{\em Example:} Independent variables: $x,y,z$, initial values:
\begin{eqnarray*}
P^x & = & A(y,z),_y + B(y,z),_z + C(y,z) + D(y) + G(z) \\
P^y & = & H(x,z),_x + K(x,z),_z + L(x) + M(x,z) + N(z) \\
P^z & = & R(x,y),_x + S(x,y),_y + T(x) + U(y) + W(x,y) \\
 & & Q^{xy} = Q^{xz} = Q^{yz} = 0
\end{eqnarray*}
containing undetermined functions $A, B, C, D, G, H, K, L, M, N, R, S,
T, U$ and $W$. After completing the program up to line 18 the values are
\begin{eqnarray*}
P^x & = &                         C(y,z) + D(y) + G(z) \\
P^y & = & H(x,z),_x             + L(x) + M(x,z) + N(z) \\
P^z & = & R(x,y),_x + S(x,y),_y + T(x) + U(y) + W(x,y) \\
Q^{xy} & = & A(y,z) \\
Q^{xz} & = & B(y,z) \\
Q^{yz} & = & K(x,z).
\end{eqnarray*}
After completing the program up to line 26 the values are
\begin{eqnarray*}
P^x & = &                         C(y,z) + D(y) + G(z) \\
P^y & = &                         L(x) + M(x,z) + N(z) \\
P^z & = &                         T(x) + U(y) + W(x,y) \\
Q^{xy} & = & A(y,z)-H(x,z) \\
Q^{xz} & = & B(y,z)-R(x,y) \\
Q^{yz} & = & K(x,z)-S(x,y).
\end{eqnarray*}
The loop beginning in line 27 will integrate the remaining terms
in $P^i$. The lines 32 - 35 will integrate the terms $D, G, L, N, T, U$
and lines 37 - 42 the terms $C, M, W$ to obtain
\begin{eqnarray*}
 & & P^x = P^y = P^z = 0 \\
Q^{xy} & = & A(y,z)-H(x,z) + y G(z) - x N(z) + F^1(y,z)  \\
Q^{xz} & = & B(y,z)-R(x,y) + z D(y) - x U(y) - F^3(x,y)  \\
Q^{yz} & = & K(x,z)-S(x,y) + z L(x) - y T(x) + F^2(x,z)
\end{eqnarray*}
with a list $F$ of new additional
functions $F^1(y,z), F^2(x,z), F^3(x,y)$ and
list $E$ of new additional equations
\begin{eqnarray*}
F^1(y,z),_y & = & C(y,z) \\
F^2(x,z),_z & = & M(x,z) \\
F^3(x,y),_x & = & W(x,y)
\end{eqnarray*}
each in less than 3 variables.

\bibliographystyle{plain}
\bibliography{syzy}

\begin{thebibliography}{10}

\bibitem{BeWe93}
T.~Becker and V.~Weispfenning.
\newblock {\em Groebner bases}.
\newblock Springer Verlag, 1993.

\bibitem{BLOP1}
F.\ Boulier, D.\ Lazard, F.~Ollivier, and M.~Petitot.
\newblock Computing representations for radicals of finitely generated
  differential ideals.
\newblock In {\em Proceedings of {ISSAC} 95}, pages 158--166. {ACM} Press,
  1995.

\bibitem{Hub1}
E.~Hubert.
\newblock Essential components of algebraic differential equations.
\newblock {\em J.\ of Symb.\ Comp.}, 28(4-5):657--680, 1999.

\bibitem{Hub2}
E.~Hubert.
\newblock Factorisation free decomposition algorithms in differential algebra.
\newblock {\em J.\ of Symb.\ Comp.}, 29(4-5), 2000.

\bibitem{KrRo00}
M.~Kreuzer and L.~Robbiano.
\newblock {\em Computational Commutative Algebra 1}.
\newblock Springer Verlag, 2000.

\bibitem{Mans}
E.L. Mansfield.
\newblock The differential algebra package diffgrob2.
\newblock {\em Mapletech}, 3:33--37, 1996.

\bibitem{Olver1986}
P.~J. Olver.
\newblock {\em Applications of {Lie} Groups to Differential Equations}, volume
  107 of {\em gtm}.
\newblock Springer Verlag, New York-Berlin-Heidelberg-Tokyo, 1986.

\bibitem{Reid2}
G.J.\ Reid, A.D.\ Wittkopf, and A.~Boulton.
\newblock Reduction of systems of nonlinear partial differential equations to
  simplified involutive forms.
\newblock {\em Europ.\ J.\ of Appl.\ Math.}, 7:604--635, 1996.

\bibitem{HPR}
T.~Wolf.
\newblock The program crack for solving {PDEs} in general relativity.
\newblock In F.W. Hehl, R.A. Puntigam, and H.~Ruder, editors, {\em Relativity
  and Scientific Computing: Computer Algebra, Numerics, Visualization}, pages
  241--251. Springer Verlag, 1996.

\bibitem{Wol99d}
T.~Wolf.
\newblock A linearization of {PDEs} based on conservation laws.
\newblock preprint, 1999.

\bibitem{Wol99e}
T.~Wolf.
\newblock The symbolic integration of exact {PDEs}.
\newblock {\em J. Symb. Comp.}, 30(5):619--629, 2000.

\bibitem{Wol99a}
T.~Wolf.
\newblock A comparison of four approaches to the calculation of conservation
  laws.
\newblock {\em Euro. Jnl of Applied Mathematics}, 13 part 2:129--152, 2002.

\bibitem{WBM}
T.~Wolf, A.~Brand, and M.~Mohammadzadeh.
\newblock Computer algebra algorithms and routines for the computation of
  conservation laws and fixing of gauge in differential expressions.
\newblock {\em J. Symb. Comp.}, 27:221--238, 1999.

\end{thebibliography}

\end{document}